\documentclass{cflowproc}

\usepackage{natbib}

\begin{document}

\newcommand\ie{{\it i.e.\ }}
\newcommand\lapprox{
        \mathrel{\lower3pt\vbox{\baselineskip=2.5pt\hbox{$<$}\hbox{$\sim$}}}}
\newcommand\gapprox{
        \mathrel{\lower3pt\vbox{\baselineskip=2.5pt\hbox{$>$}\hbox{$\sim$}}}}


\shortauthors{NULSEN}     
\shorttitle{AGN INTERACTION} 

\title{Interaction of the AGN and X-ray Emitting Gas}   

\author{Paul Nulsen\affilmark{1,2}}   

\affil{1}{University of Wollongong, Wollongong NSW 2522, Australia}
\and
\affil{2}{Present address: Harvard-Smithsonian Center for
Astrophysics, 60 Garden St, Cambridge MA 02138}                                


\begin{abstract}
The process that prevents the deposition of cooled gas in cooling
flows must rely on feedback in order to maintain gas with short
cooling times, while preventing the bulk of the gas from cooling to
low temperatures.  The primary candidate for the feedback mechanism is
the accretion of cooled and cooling gas by an active galactic nucleus
(AGN).  Despite some difficulties with this model, the high incidence
of central radio sources in cooling flows and the common occurrence of
radio lobe cavities, together, support this view.  The Bondi accretion
rate for the intracluster gas onto the AGN depends on the gas
properties only through its specific entropy and that is governed
directly by competition between heating and cooling.  This provides a
viable link for the feedback process.  It is argued that the mass
accreted between outbursts by the central AGN is only sensitive to the
mass of the black hole and the gas temperature.  Bondi accretion by an
AGN leads to a simple expression for outburst energy that can be
tested against observations.
\end{abstract}


\section{Introduction --- the Significance of AGN Heating}
\label{Nulsen:intro}

A variety of mechanisms have been proposed to explain why large
quantities of gas do not cool to low temperatures at the centers of
cooling flows (many discussed elsewhere in these proceedings).  While
there are many candidate heating mechanisms, in general, short central
cooling times, as little as $10^8$ -- $10^9$ y in many clusters
\citep[e.g.][]{dnm,tsb} and an order of magnitude shorter still in
elliptical galaxies \citep[e.g.][]{mb03}, are very difficult to
maintain without a process involving feedback.  In the absence of
feedback, any heating process will either overwhelm, or be overwhelmed
by radiative cooling in a few cooling times.  Adding to this the
observation that outbursts from central AGN's have an impact on the
the hot gas through the creation of cavities around radio lobes
\citep[e.g.][]{mwn,fse,bsm} in clusters, groups and isolated
elliptical galaxies, and the prevalence of radio sources at the
centers of cooling flows \citep{burns}, suggests that AGN feedback is
the prime candidate as the process that prevents mass deposition in
cooling flows. 

This is not to dismiss the difficulties faced by the AGN feedback
model \citep[e.g.][and elsewhere in these proceedings]{fmn,bm03}.
However, it has been well demonstrated that a modest amount of thermal
conduction could make a significant contribution to the energy budget
of the regions in clusters where the cooling time is shorter than the
age.  Thermal conduction could, therefore, alleviate the major problem
of the AGN feedback model in clusters, but as also shown by
\citet{zn}, thermal conduction cannot prevent the deposition of cooled
gas in all cluster cooling flows.  The steep temperature dependence of
the thermal conductivity ($\kappa \sim T^{5/2}$) means that thermal
conduction is less effective in groups, and less still in isolated
elliptical galaxies.  Furthermore, the region in clusters where the
cooling time is shorter than the Hubble time is relatively small, so
that the remainder of the cluster can act as a thermal reservoir to
make up for radiative losses from the center.  This is not the case in
isolated elliptical galaxiess, where radiative losses are significant
throughout the system.  Thus, while thermal conduction may be
important for the energy budget of cluster cooling flows, it is
unlikely to have much impact in cooler systems.

For the clusters with radio cavities studied to date, single AGN
outbursts have very little impact on the cluster as a whole.  However,
the opposite is true for isolated elliptical galaxies.  This is clear
from observations of systems such as M84 \citep{fj}, where a single
radio outburst has disrupted the entire hot interstellar medium of the
galaxy.  Thus, while there is room to doubt the significance of
heating by AGN outbursts in cluster cooling flows, there is a strong
case that they can regulate the deposition of cooled gas in elliptical
galaxies \citep[but see][]{bm03}.  It would be surprising if the
mechanism that regulates cooling in the larger systems was
fundamentally different.

If AGN feedback is the process that regulates the deposition of cooled
gas in cooling flows, then it must be powered by the accretion of
cooled and cooling gas onto the AGN.  The remainder of my talk is a
simpleminded consideration of what we should expect in that case.

\section{Bondi Acretion from Cooling Hot Gas}

It is well known that angular momentum plays a critical role in
accretion onto a black hole, \ie an AGN.  Since the ratio of the Bondi
radius, where the influence of the black hole on the surrounding gas
starts to be significant, to the Schwarzchild radius is roughly
$c^2 /s^2 \sim 10^5$ -- $10^6$, angular momentum that is negligible at
the Bondi radius can have a dominant effect on the subsequent flow.
Here it is assumed that accreted gas is fed into a reasonably compact
accretion disc, from where it is fed into the AGN.  At the least, this
means that there need not be a very close link between the Bondi
accretion rate and the instantaneous accretion rate onto the AGN.
Nevertheless, gas that is fed into the disc is assumed to be available
to fuel the AGN. It should also be noted that, based on numerical
simulations, \citet{pb03a,pb03b} have cast considerable doubt on the
Bondi accretion formula.  Despite this, it is employed here for want
of a better replacement.  It is unclear from the results of
\citet{pb03b} under what conditions, if any, the Bondi accretion rate
may apply.  In particular, those simulations are adiabatic and when
radiative cooling becomes appreciable at the Bondi radius ($\beta \to
1$ below) the Bondi result may be more accurate, and the estimate of
the accreted mass would still apply.

The ratio of specific heats in the intracluster gas is $\gamma=5/3$,
so that the Bondi accretion rate depends on the gas properties through
its entropy alone.  The accretion rate from uniform gas with density
$\rho_0$ and temperature $T_0$ may be expressed as
\begin{equation}\label{bondiacc}
\dot M_{\rm B} = \pi {(GM_{\rm h})^2 \over \Sigma_0^{3/2}}
\simeq 0.012 n_{\rm e} T^{-3/2} M_9^2 \rm\ M_\odot\ y^{-1},
\end{equation}
where the black hole mass is $M_{\rm h} = 10^9M_9 \rm\ M_\odot$, the
entropy is represented by $\Sigma_0 = s_0^2\rho_0^{-2/3}$, with the
sound speed $s_0 = 519 T^{1/2}\rm\  km\ s^{-1}$, the electron density
corresponding to $\rho_0$ is $n_{\rm e}\rm\ cm^{-3}$ and $T$ is $kT_0$
in keV.  It should also be noted that for $\gamma=5/3$ the sonic point
in the accretion flow occurs at $R=0$, and the accretion rate is
determined by the entropy of the gas at this point.  As a result,
the galactic environment has little effect on the Bondi accretion rate
(and we can insert the density and temperature at any point in the
flow into equation (\ref{bondiacc}) to get the accretion rate).

Radiative cooling directly affects the entropy of the gas, hence the
Bondi accretion rate.  Expressed in terms of the specific entropy $S$,
the energy equation for the gas may be written
\begin{equation}\label{entropy}
\rho T {dS\over dt} = - n_{\rm e} n_{\rm H} \Lambda(T),
\end{equation}
where $d/dt$ is the convective or lagrangian time derivative.  This
equation can be rewritten in terms of the entropy proxy, $\Sigma$
above, as 
\begin{equation}\label{entproxy}
{d\ln\Sigma\over dt} = -{1\over t_{\rm cool}},
\end{equation}
where the radiative cooling time is defined as usual,
\begin{equation}\label{tcool}
t_{\rm cool} = {3 \rho k T \over 2 \mu m_{\rm H} n_{\rm e} n_{\rm H}
\Lambda(T)}.
\end{equation}
This makes the link between cooling and Bondi accretion very clear.
Cooling reduces the entropy, as in equation (\ref{entproxy}), and the
entropy determines the Bondi accretion rate, as in equation
(\ref{bondiacc}).  Any heat input to the gas that results from AGN
activity would be added to the right hand side of equation
(\ref{entropy}), allowing us to close the feedback loop.  Of course,
the details of this process may be complex and, in particular, cycles
of heating, accretion and cooling may be intermittent rather than
steady.

The behavior of cooling gas is relatively simple.  Radiative losses
reduce the entropy of the gas, which is therefore compressed by
surrounding gas in order to maintain hydrostatic equilibrium.  This
causes inflow.  The most counterintuitive feature of this flow is
that, as a result of the compression, the temperature of the cooling
gas is maintained at about the ``virial temperature'' of the
gravitational potential in which it resides.  That is, the gas
temperature at radius $R$ is given by $kT(R)/(\mu m_{\rm H}) \simeq
GM(R)/R$ (to within a factor of order unity), where $M(R)$ is the
gravitating mass within $R$ (this only fails if the virial temperature
decreases too rapidly with decreasing radius).

The primary condition required to maintain the temperature of the
cooling gas close to the virial temperature is that the gas must
remain approximately hydrostatic.  The condition for this is that the
cooling time of the gas exceeds its sound crossing time, \ie at radius
$R$, that $t_{\rm cool} \gapprox t_{\rm sc} = R/s$, where $s$ is the
speed of sound in the gas.  If the cooling time falls below the sound
crossing time, the gas will cool to low temperature, more-or-less in
place (\ie isochorically), and its subsequent behavior is strongly
time-dependent.

For the Bondi accretion flow, what matters is the ratio of cooling
time to sound crossing time at about the Bondi radius.  Inside the
Bondi radius, the ratio of cooling time to sound crossing time
increases inward in the Bondi flow ($\propto T/\Lambda(T)$), so that
radiative cooling is unimportant throughout if it is unimportant at
the Bondi radius.  The ratio of cooling time to sound crossing time at
the Bondi radius is
\begin{equation}\label{ratbondi}
\beta = \left. t_{\rm cool} \over t_{\rm sc} \right|_{r_{\rm B}}
\simeq {2 kT_0 s_0^3 \over \mu n_{\rm e, 0} GM_{\rm h} \Lambda(T_0)}
\end{equation}
(where we have used $\rho_0/n_{\rm H,0} = 4 m_{\rm H} / 3$, as
appropriate for gas that is 25 percent helium by mass).  Thus, as the
gas cools, the temperature of the gas at the Bondi radius remains
nearly constant, at about the virial temperature as long as $\beta >
1$.  If the cooling time becomes shorter than the sound crossing time
at the Bondi radius, then the gas temperature at the Bondi radius will
plummet.  Formally, this causes a dramatic increase in the Bondi
radius and accretion rate, but in practice the flow will become
strongly time-dependent.  Gas near to the Bondi radius will go into
free-fall, and we should expect a sharp rise in the accretion rate
onto the AGN.

It is instructive to express the Bondi accretion rate in terms of
$\beta$.  Using equation (\ref{ratbondi}) to determine the electron
density, the ratio of the Bondi accretion rate to the ``Eddington''
accretion rate is
\begin{equation}
\dot m_{\rm B} = {\dot M_{\rm B} \over \dot M_{\rm Edd}}
\simeq {\eta \sigma_{\rm T} c k T_0 \over \beta\Lambda(T_0)}
\simeq 0.3 {\eta_{-1} T \over \beta \Lambda_{-23}},
\end{equation}
where $\Lambda(T_0) = 10^{-23} \Lambda_{-23} \rm\ erg\ cm^3\ s^{-1}$,
and the ``Eddington'' accretion rate is defined as usual by
\begin{equation}
\eta \dot M_{\rm Edd} c^2 = L_{\rm Edd} = {4\pi GM_{\rm h} m_{\rm H} c
\over \sigma_{\rm T}},
\end{equation}
where $L_{\rm Edd}$ is the Eddington luminosity and $\eta = 0.1
\eta_{-1}$ is the radiative efficiency of accretion onto the AGN.  The
significant feature to note is that the Bondi accretion rate from the
hot gas approaches the Eddington accretion rate as $\beta$ approaches
1.

\section{Outburst Energy}

In what follows, it is assumed that, in the absence of feedback from
the AGN, the entropy of the gas arriving at the Bondi radius decreases
with time due to cooling.  However, if the Bondi accretion rate always
exceeds the rate at which gas would cool to low temperatures in a
simple cooling flow, then Bondi accretion can outcompete cooling and
stop the entropy of the gas at the Bondi radius from decreasing.  At
least for the case of groups and clusters of galaxies, where the old
(morphological) estimates of cooling rate greatly exceed the Bondi
accretion rate \citep[e.g.][]{fabian}, this cannot occur.

While it is unclear what would bring on an AGN outburst, it seems
very likely that it is related to the accretion rate.  For example, it
has been argued for some time that the radiative efficiency of AGN
accretion switches from low to high when the accretion rate exceeds a
small fraction of the Eddington accretion rate, $\dot M_{\rm Edd}$
\citep{nar,abr}.  This may also lead to a sudden increase in the
mechanical energy output of the AGN.  Alternatively, the accretion
rate may need to reach closer to $\dot M_{\rm Edd}$ or even exceed it
somewhat to cause an outburst.  For the purpose of the argument that
follows, the only significant issue is that an outburst occurs when
the accretion rate reaches some multiple of the Eddington accretion
rate that is not a lot more than unity.  In that case, we may assume
that the gas is approximately hydrostatic at the Bondi radius up to
the onset of the outburst.

We can now calculate the mass of gas accreted before the outburst, or,
more precisely, the mass of fuel available to power an outburst.  The
mass accreted in the time interval $t_{\rm i}$ to $t_{\rm f}$ is
\begin{equation}
M_{\rm a} = \int_{t_{\rm i}}^{t_{\rm f}} \dot M_{\rm B} \, dt
\simeq \int_{\Sigma_{\rm f}}^{\Sigma_{\rm i}} \dot M_{\rm B} t_{\rm
cool} \, {d\Sigma_0 \over \Sigma_0},
\end{equation}
where equation (\ref{entproxy}) has been used to change variable in
the integral, and $\Sigma_{\rm i}$ and $\Sigma_{\rm f}$ are the
entropies of the accreting gas at $t_{\rm i}$ and $t_{\rm f}$,
respectively.  This change of variable involves an approximation,
since equation (\ref{entproxy}) applies to a single fluid element,
whereas different fluid elements are continually being accreted.  The
approximation relies on the gas entering the accretion flow being
nearly uniform.  In general, it is fairly crude, but
we are assisted by two things.  First, it is likely that the accreted
gas that fuels the outburst comes, initially, from a fairly narrow
range of radii (in fractional terms), so that it is likely to be
reasonably uniform.  Second, the sense of our error is to
underestimate $dt$, hence the total mass accreted.  This is because
the entropy is a non-decreasing function of the radius in a stably
stratified atmosphere and the pressure is a decreasing function of the
radius.  Both effects mean that cooling is slower at larger radii, so
that the decreasing entropy of an outer gas shell lags behind that
for an inner one.

Using the Bondi accretion rate (\ref{bondiacc}) and the cooling time
(\ref{tcool}), we find that $\dot M_{\rm B} t_{\rm cool}$ only depends
on the gas temperature.  As argued above, this remains constant until
the cooling time is comparable to the sound crossing time at the Bondi
radius.  Thus the integral gives
\begin{equation}\label{accreted}
M_{\rm a} \simeq \dot M_{\rm B, i} t_{\rm cool, i} \ln{\Sigma_{\rm i}
\over \Sigma_{\rm f}}.
\end{equation}
While we have only a vague idea of the values to use for $\Sigma_{\rm
i}$ and $\Sigma_{\rm f}$, the result is quite insensitive to these.
We therefore take $M_{\rm a} = \chi \dot M_{\rm B,i} t_{\rm cool,i}$,
where the latter two factors can be evaluated at any one time when the
cooling time exceeds the sounding crossing time at the Bondi radius.
Allowing a reasonable range of $\Sigma$ in the logarithmic factor, and
including some correction for the crude approximation involved in the
change of variable from $t$ to $\Sigma$, reasonable values of $\chi$
are of the order of 10.

Using the results from above, we get (again assuming 25 percent helium
by mass)
\begin{equation}
M_{\rm a} \simeq {48 \pi \chi (GM_{\rm h} m_{\rm H})^2 \over 35 s_0
\Lambda(T_0)}
\simeq 2\times 10^5 {\chi M_9^2 \over T^{1/2} \Lambda_{-23}} \rm \
M_\odot,
\end{equation}
using the notation from above for the numerical factors.  If this mass
is accreted onto the black hole and converted to mechanical energy
with efficiency $\eta_{\rm mech}$, then the energy released is
\begin{equation}\label{outburst}
E_{\rm a} = \eta_{\rm mech} M_{\rm a} c^2
\simeq 3.7\times10^{58} {\eta_{\rm m, -1} \chi M_9^2 \over T^{1/2}
\Lambda_{-23}} \rm\ erg,
\end{equation}
where $\eta_{\rm mech} = 0.1 \eta_{\rm m, -1}$.  This is our estimate
of the energy available to power an outburst.

\section{Application to Hydra A}

One of the key assumptions made in the calculation above is that the
feedback occurs in outbursts.  This is, at least, consistent with the
observation that not all cooling flows contain an active central radio
source \citep{burns}.  If the cooling flows that lack radio sources
are in between outbursts, then this also tells us that the evidence of
outbursts fades with time.  Indeed, the radio lobe cavities, which are
the main evidence of mechanical energy input, almost certainly rise
buoyantly and disappear, either because the decreasing contrast of the
rising cavity makes it difficult to detect, or because the bubbles are
non-adiabatic, so that their energy is dissipated and the bubbles
collapse.  As a result, estimates of outburst energy based on
observations of bubbles can only give lower limits to actual outburst
energies.

Hydra A is noteworthy for being one of the most powerful
Fanaroff-Riley class I radio sources and for having substantial
bubbles associated with its radio lobes \citep{mwn}.  Using the gas
properties from \citet{dnm}, the central temperature is $kT\simeq 3$
keV and the central abundance is $Z\simeq 1$, giving the cooling
function $\Lambda_{-23} \simeq 1.8$.  \citet{sambruna} estimated the
mass of the black hole in Hydra A as $M_h \simeq 4 \times 10^9\rm\
M_\odot$.  Using these numbers in equation (\ref{outburst}) gives an
outburst energy of $E_{\rm a} \simeq 2 \times 10^{59} \chi \eta_{\rm
m, -1}$ erg. 

For the cavity created by the southwest radio lobe in Hydra A, the
work required to inflate the cavity is $p V \simeq 2 \times 10^{59}$
erg.  This must be added to the thermal energy in the lobe, $p V /
(\gamma -1)$, where $\gamma$ is the ratio of specific heats for the
plasma inside the cavity, to obtain the total energy required to
create the lobe.  If the lobe plasma is relativistic, $\gamma=4/3$ and
the energy required to create the lobe is $4 pV$.  This quantity needs
to be doubled again to allow for the northeast cavity, giving a total
energy for the bubbles of about $1.6\times10^{60}$ erg.  This agrees
with the esimate of outburst energy for the model if $\chi\eta_{\rm m,
-1} \simeq 8$.  Note that, if the expanding cavities drove shocks into
the intergalactic medium, then even more energy was required to create
them.

Given the uncertainties, these results are broadly consistent with the
outburst model described here.  Clearly, a lot more data are required
to test this model.  As discussed above, we should expect to find
evidence of a range of outburst energies, extending up to the value
given by equation (\ref{outburst}).  A more detailed model for the
evolution of the bubbles is required to determine the distribution of
bubble energies we should expect to observe, but we can reasonably
expect bubbles to evolve on timescales comparable to their rise times
\citep{chur}.  Since these are comparable to the estimated intervals
between radio outbursts, a significant fraction of systems are
expected to show bubble energies comparable to the outburst energy.
Of course, outbursts of significantly greater energy than given by
equation (\ref{outburst}) could not be accounted for by this model.

The outburst energy in Hydra A is close to the maximum that can be
accounted for using (\ref{outburst}).  If the accretion rate remained
an order of magnitude or more smaller than the Bondi accretion rate
throughout the time between outbursts \citep{pb03b}, then the AGN
could not accrete enough hot gas to power formation of the bubbles in
Hydra A.  In that case, the AGN would need to be powered chiefly by
accretion of cold gas.  This result is not sensitive to the model
assumptions used here.

\section{Conclusions}

Heating by AGN feedback may not be the only mechanism that prevents
the deposition of cool gas in cooling flows, but it almost certainly
plays a significant role.  AGN heating may be augmented by other
processes, particularly thermal conduction in clusters, but it could
be the only heating mechanism required in smaller systems, expecially
for isolated elliptical galaxies.  This is not to dismiss the major
gaps in our understanding of the heating process
\citep[e.g.][]{fmn,bm03}.

Because radiative cooling decreases gas entropy, directly increasing
the Bondi accretion rate, Bondi accretion of cooling gas onto a
central black hole makes a good candidate for part of the AGN feedback
process.  Here, Bondi accretion is assumed to provide the fuel for AGN
outbursts.  The Bondi accretion rate climbs close to the Eddington
accretion rate as the cooling time at the Bondi radius falls towards
the sound crossing time.  An outburst can be triggered at about this
stage, perhaps when the radiative efficiency of accretion switches
from low to high, or later when the luminosity reaches close to the
Eddington limit.  Accreting gas remains roughly hydrostatic near
to the Bondi radius as long as the cooling there is shorter than the
sound crossing time.  In that case the temperature of gas at the Bondi
radius stays close to the virial temperature as the AGN accretes
fuel for the outburst.  The mass of fuel accreted is then insensitive
to the details of when the outburst occurs.  This gives an estimate
for outburst energy, equation \ref{outburst}, which depends weakly on
uncertain details.  This outburst energy is determined mostly by the
mass of the black hole and the virial temperature of its host galaxy.



\acknowledgements

This work was partly supported by the Australia Research Council and
by NASA grant NAS8-01130.


\end{document}